\documentclass[prd,12pt,nofootinbib,superscriptaddress]{revtex4-2}
\usepackage{bm}
\usepackage{amsmath}
\usepackage{amssymb}
\usepackage{booktabs}
\usepackage{graphicx}
\usepackage{float}
\usepackage{longtable}
\usepackage{subfigure}
\usepackage{hyperref}
\usepackage{comment}
\usepackage{mathrsfs}
\usepackage{xcolor}
\usepackage{color}
\usepackage{ulem}
\usepackage{CJK}
\hypersetup{colorlinks=true,linkcolor=red,citecolor=blue}

\begin{document}
\title{On the duality in constant-roll inflation}
\author{Yue Wang}
\email{ywang123@hust.edu.cn}
\affiliation{School of Physics, Huazhong University of Science and Technology, Wuhan, Hubei
430074, China}
\author{Qing Gao}
\email{Corresponding author. gaoqing1024@swu.edu.cn}
\affiliation{School of Physical Science and Technology, Southwest University,
Chongqing 400715, China}
\author{Shengqing Gao}
\email{gaoshengqing@hust.edu.cn}
\affiliation{School of Physics, Huazhong University of Science and Technology,
Wuhan, Hubei
430074, China}
\author{Yungui Gong}
\email{gongyungui@nbu.edu.cn}
\affiliation{Department of Physics, School of Physical Science and Technology, Ningbo University, Ningbo, Zhejiang 315211, China}
\affiliation{School of Physics, Huazhong University of Science and Technology, Wuhan, Hubei
430074, China}
\begin{abstract}
There is a duality in the observables $n_s$, $r$ and the inflaton potential between large and small $\eta_H$ for the constant-roll inflation if the slow-roll parameter $\epsilon_H$ is negligible.
In general, the duality between $\eta_H$ and $\bar{\eta}_H$ does not hold for the background evolution of the inflaton. 
For some particular solutions for the constant-roll inflation with $\eta_H$ being a constant, 
we find that in the small field approximation, 
the potential takes the quadratic form and it remains the same when the parameter $\eta_H$ changes to $\bar{\eta}_H=3-\eta_H$.
If the scalar field is small and the contribution of $\epsilon_H$ is negligible, 
we find that there exists the logarithmic duality and the duality between large and small $\eta_H$ for the primordial curvature perturbation in inflationary models with the quadratic potential.

\end{abstract}

\maketitle

\section{Introduction}
As more evidences for the existence of 
primordial black holes (PBHs) were provided by the observations of gravitational waves (GWs) such as the 
Laser Interferometer Gravitational Wave Observatory (LIGO) Scientific and Virgo Collaborations \cite{Bird:2016dcv,Sasaki:2016jop,Inomata:2016rbd,DeLuca:2020sae,DeLuca:2021wjr,Franciolini:2021tla},
and the Pulsar Timing Arrays (PTAs)
\cite{DeLuca:2020agl, Vaskonen:2020lbd, NANOGrav:2023hvm,EPTA:2023xxk},
PBHs as dark matter attracted a lot of attention. 
When the density contrast of overdense regions exceeds
the threshold value at the horizon reentry during radiation
domination,
PBHs may form  through gravitational collapse \cite{Carr:1974nx,Hawking:1971ei}. 
The large density contrast of overdense regions may be seeded from large primordial
curvature perturbations at small scales generated during inflation. The large curvature perturbations are also the sources of secondary gravitational waves (GWs) after the horizon reentry through the scalar-tensor mixing \cite{Matarrese:1997ay,Mollerach:2003nq,Ananda:2006af,Baumann:2007zm}.
Therefore, accompanied with the production of PBHs, scalar induced gravitational waves (SIGWs) are generated  \cite{Matarrese:1997ay,Mollerach:2003nq,Ananda:2006af,Baumann:2007zm,Garcia-Bellido:2017aan,Saito:2008jc,Saito:2009jt,Bugaev:2009zh,Bugaev:2010bb,Alabidi:2012ex,Inomata:2019ivs,Braglia:2020taf,Yi:2022anu,Orlofsky:2016vbd,Nakama:2016gzw,Inomata:2016rbd,Cheng:2018yyr,Cai:2018dig,Bartolo:2018rku,Bartolo:2018evs,Kohri:2018awv,Espinosa:2018eve,Cai:2019amo,Cai:2019elf,Cai:2019bmk,Cai:2020fnq,Domenech:2019quo,Domenech:2020kqm,Pi:2020otn,Germani:2017bcs, Ezquiaga:2017fvi, Ballesteros:2017fsr,Gao:2018pvq, Dalianis:2019vit, Ballesteros:2018wlw,Lu:2019sti,  Passaglia:2018ixg, Fu:2019ttf, Lin:2020goi, Yi:2020kmq, Yi:2020cut, Wang:2024euw}.

To produce large curvature perturbations at small scales during inflation,
we usually consider inflationary models with an inflection point in
the inflaton potential \cite{Di:2017ndc,Espinosa:2017sgp, Garcia-Bellido:2017mdw,Sasaki:2018dmp}.
Near the inflection point, the slow-roll parameter $\eta_H\approx 3$ and it is called the ultra-slow-roll (USR) inflation \cite{Tsamis:2003px,Kinney:2005vj}. 
More general, the constant-roll inflation with $\eta_H$ being a constant was proposed \cite{Martin:2012pe,Motohashi:2014ppa}.
If $\eta_H>1$, then the slow-roll condition is violated in constant-roll inflation and the primordial curvature perturbations may evolve outside the horizon \cite{Leach:2000yw,Leach:2001zf,Kinney:2005vj,Jain:2007au,Namjoo:2012aa,Martin:2012pe,Motohashi:2014ppa,Yi:2017mxs}.  
For constant-roll inflation, the inflationary potential and the background equation of motion can be solved analytically.
Due to different background evolution for large and small $\eta_H$,
the observational data constrained $\eta_H$ to be small \cite{Motohashi:2017aob,Gao:2018cpp,GalvezGhersi:2018haa,Gao:2019sbz}.
Since the slow-roll parameter $\epsilon_H$ decreases with time,
so it may be negligible during inflation.
By neglecting the contribution from $\epsilon_H$, it was found that there exists a duality between $\eta_H$ and $\bar{\eta}_H=3-\eta_H$ in the observables the scalar spectral tilt $n_s$ and the tensor-to-scalar ratio $r$ \cite{Tzirakis:2007bf,Morse:2018kda}.
The duality between $\eta_H$ and $\bar{\eta}_{H}=3-\eta_H$ connects the constant slow-roll inflation with $|\eta_H|\ll 1$ and the USR inflation with $\eta_H\approx 3$,
so the behavior of the primordial curvature perturbations in USR inflation can be understood from the usual slow-roll inflation.
Recently, the logarithmic duality of the primordial curvature perturbation was found for the quadratic potential in Ref. \cite{Pi:2022ysn}. 
Since the logarithmic duality applies to both the slow-roll and USR inflation, 
we extend the discussion on the duality to constant-roll inflation in this paper.

The paper is organized as follows.
In Section \ref{sec2}, we discuss the duality in constant-roll inflation.
Motivated by the duality in constant-roll inflation, we then discuss the logarithmic duality of the primordial curvature perturbation using the $\delta N$ formalism in Section \ref{sec3}.
The conclusion is drawn in Section \ref{sec4}.

\section{The constant-roll inflation}\label{sec2}
For the constant-roll inflation, we take the second Hubble flow slow-roll parameter
\begin{equation}
\label{etaconstant1}
\eta_{H}=\frac{2}{H}\frac{d^2H}{d\phi^2}=-\frac{\ddot{H}}{2H\dot{H}}=-\frac{\ddot{\phi}}{H\dot{\phi}}
\end{equation} 
to be a constant, i.e., we take $\eta_{H}=3+\alpha$ with $\alpha$ being a constant.
If $\alpha\approx -3$, then $\eta_H$ is small and the constant-roll inflation is also a slow-roll inflation; 
If $\alpha\approx 0$, then $\eta_H\approx 3$ and $dV/d\phi\approx 0$,
the constant-roll inflation is USR inflation. 
For the constant-roll inflation with a constant $\eta_H$, the scalar spectral tilt is \cite{Motohashi:2014ppa,Yi:2017mxs}
\begin{equation}
\label{nseq1}
n_s-1\approx 3-|2\eta_H-3|-\frac{2(2\eta_H^2-9\eta_H+6)}
{|2\eta_H-3|}\epsilon_H,
\end{equation}
and the tensor-to-scalar ratio is \cite{Motohashi:2014ppa,Yi:2017mxs}
\begin{equation}
\label{ra}
r\approx2^{7-|2\eta_H-3|}\left(\frac{\Gamma[3/2]}{\Gamma[|2\eta_H-3|/2]}\right)^2 \epsilon_H,
\end{equation}
where  $\epsilon_{H}=2(dH/d\phi)^{2}/H^{2}=\dot{\phi}^2/(2H^2)$ is the first Hubble flow slow-roll parameter.
If the contribution of $\epsilon_H$ is negligible in Eq. \eqref{nseq1},
then the scalar spectral tilt remains unchanged with the replacement of $\eta_H$ by $\bar{\eta}_H=3-\eta_H$.
It is interesting to note that the tensor-to-scalar ratio also keeps the same under the interchange between 
$\eta_H$ and $\bar{\eta}_H$.
The behaviours of $n_s$ and $r$ under the swap between $\eta_H$ and $\bar{\eta}_H=-\alpha$ in the constant-roll inflation
are called the duality between large and small $\eta_H$ in the observables $n_s$ and $r$ \cite{Tzirakis:2007bf,Morse:2018kda}.
The duality connects the slow-roll inflation with $|\eta_H|\ll 1$ and the USR inflation with $\eta_H\approx 3$,
so it is useful for the understanding of PBH formation and SIGW generation by USR inflation from slow-roll inflation. 
If the contribution of $\epsilon_H$ in Eq. \eqref{nseq1} is not negligible,
then the duality in $n_s$ does not exist, and the background evolutions are also very different \cite{Gao:2019sbz}.
Therefore, it is important to further explore the duality between $\eta_H$ and $\bar{\eta}_H$. 

To see when $\epsilon_H$ can be neglected, 
substituting $\eta_H=3+\alpha$ into Eq. \eqref{nseq1}, we get
\begin{equation}
\label{nseq2}
n_s-1\approx -2\alpha-\frac{2(2\alpha^2+3\alpha-3)}
{3+2\alpha}\epsilon_H,
\end{equation}
where $\alpha>-3/2$. Therefore, the condition that the contribution of $\epsilon_H$ is negligible is
\begin{equation}
\label{negcond1}
\epsilon_H\ll \left|\frac{\alpha(3+2\alpha)}{2\alpha^2+3\alpha-3}\right|.
\end{equation}
For the USR inflation, $|\alpha|\ll 1$, the condition \eqref{negcond1} means that $\epsilon_H\ll |\alpha|\ll 1$.
Substituting $\bar{\eta}_H=-\alpha$ into Eq. \eqref{nseq1}, we get
\begin{equation}
\label{nseq3}
n_s-1\approx -2\alpha-\frac{2(2\alpha^2-9\alpha+6)}
{3+2\alpha}\epsilon_H.
\end{equation}
For the case with $\bar{\eta}_H$, the condition that the contribution of $\epsilon_H$ is negligible is
\begin{equation}
\label{negcond2}
\epsilon_H\ll \left|\frac{\alpha(3+2\alpha)}{2\alpha^2-9\alpha+6}\right|.
\end{equation}
For slow-roll constant inflation with $|\alpha|\ll 1$, the condition \eqref{negcond2} becomes $\epsilon_H\ll |\alpha|/2\ll 1$.
Combining the results \eqref{negcond1} and \eqref{negcond2}, we see that
the duality between the USR inflation with $\eta_H=3+\alpha$ and the slow-roll constant inflation with $\bar{\eta}_H=-\alpha$ holds when $\epsilon_H\ll |\alpha|/2\ll 1$.

Now we discuss the background evolution of the constant-roll inflation with negligible $\epsilon_H$.
From the background equation
\begin{equation}
\label{friedman} 
3H^{2}=\frac{1}{2}\dot{\phi}^{2}+V(\phi), 
\end{equation}
we get 
\begin{equation}
\label{Vphi}
V(\phi)=(3-\epsilon_H)H^2,
\end{equation}
and
\begin{equation}
\label{dvdphi}
\frac{dV}{d\phi}=(3-\eta_{H})H^2\frac{d\phi}{dN}.
\end{equation}

In terms of the number of e-folds $N$, $dN=-Hdt$, the equation of motion for the inflaton becomes
\begin{equation}
\label{eomphi}
\frac{d^{2}\phi}{dN^{2}}+(\epsilon_{H}-3)\frac{d\phi}{dN}+\frac{1}{H^{2}}\frac{dV}{d\phi}=0.
\end{equation}
Combining Eqs. \eqref{dvdphi} and \eqref{eomphi},
we get 
\begin{equation}
\label{EOMN}
\frac{d^{2}\phi}{dN^{2}}+(\epsilon_H-\eta_{H})\frac{d\phi}{dN}=0.
\end{equation}
Eq. \eqref{EOMN} is just the definition \eqref{etaconstant1} of $\eta_H$.
If $\epsilon_{H}$ is negligible, then Eqs. \eqref{Vphi}, \eqref{dvdphi}  and \eqref{EOMN} become
\begin{gather}
\label{Vphi1}
V(\phi) \approx 3H^{2}(\phi),\\
\label{dvdphi2}
\frac{dV}{d\phi}\approx \frac{3-\eta_H}{3}V(\phi)\frac{d\phi}{dN},\\
\label{EOMN1}
\frac{d^{2}\phi}{dN^{2}}-\eta_{H}\frac{d\phi}{dN}\approx 0.
\end{gather}
The solution to Eq. \eqref{EOMN1} is
\begin{equation}
\label{phin1}
\phi(N)\approx A e^{\eta_H N}+B,
\end{equation}
where $A$ and $B$ are integration constants.
Substituting the solution \eqref{phin1} into Eq. \eqref{dvdphi2}, we get
\begin{equation}
\label{dvdphi3}    
\frac{dV}{V}\approx \frac{1}{3}\eta_H(3-\eta_H)(\phi-B)d\phi.
\end{equation}
So the potential is
\begin{equation}
\label{vphisol3}  
V(\phi)\approx V_0\exp\left[\frac{1}{3}\eta_H(3-\eta_H)\left(\frac{1}{2}\phi^2-B\phi\right)\right],
\end{equation}
and $H^2\approx V(\phi)/3$, where $V_0$ is an integration constant. 
Therefore, if $\epsilon_H$ is negligible, then the potential \eqref{vphisol3} and the Hubble parameter $H(\phi)$ are unchanged under the transformation from $\eta_H$ to $\bar{\eta}_H=3-\eta_H$, 
i.e., both $H(\phi)$ and $V(\phi)$ have the duality between $\eta_H$ and $\bar{\eta}_H$.

\section{Logarithmic duality of primordial curvature perturbation}\label{sec3}

From the definition \eqref{etaconstant1} of $\eta_H$, we get
\begin{equation}
\label{heq1}
\frac{d^{2}H}{d\phi^{2}}-\frac{\eta_H}{2}H=0.
\end{equation}
The solution to Eq. \eqref{heq1} is
\begin{equation}
\label{Hphi}
H(\phi)=C_{1}\exp\left(\sqrt{\frac{\eta_H}{2}}\,\phi\right)+C_{2}\exp\left(-\sqrt{\frac{\eta_H}{2}}\,\phi\right),
\end{equation}
for $\eta_H>0$,
and
\begin{equation}
\label{Hphi1}
H(\phi)=C_{1}\cos\left(\sqrt{-\frac{\eta_H}{2}}\,\phi\right)+C_{2}\sin\left(\sqrt{-\frac{\eta_H}{2}}\,\phi\right),
\end{equation}
for $\eta_H<0$,
where $C_{1}$ and $C_{2}$ are integration constants. 
Substituting the solutions \eqref{Hphi} and \eqref{Hphi1} into Eq. \eqref{Vphi}, we can obtain the potential $V(\phi)$. 
Therefore, the potential and the background evolution of the inflaton are determined by the parameter $\eta_H$ only for the constant-roll inflation with the slow-roll parameter $\eta_H$ being a constant \cite{Motohashi:2014ppa}.

For the particular solution 
\begin{equation}
\label{hphib}
H(\phi)=M\sinh\left(\sqrt{\frac{\eta_{H}}{2}}\phi\right),
\end{equation}
the potential is
\begin{equation}
\label{vphib}
V(\phi)=M^2\left[3\sinh^2\left(\sqrt{\frac{\eta_H}{2}}\phi\right)-\eta_H\cosh^2\left(\sqrt{\frac{\eta_H}{2}}\phi\right)\right]. 
\end{equation}
In the small field approximation with $|\phi|\ll 1$,
we get
\begin{equation}
\label{vphibapprox}
V(\phi)\approx M^2\left[-\eta_H+\eta_H(3-\eta_H)\frac{\phi^2}{2}\right]. 
\end{equation}
Because of the presence of the first term $-\eta_H$ in the right hand side of Eq. \eqref{vphibapprox}, 
the duality between $\eta_H<3$ and $\bar{\eta}_H=3-\eta_H$ 
does not hold in both $H(\phi)$ and $V(\phi)$ for the solutions \eqref{hphib} and \eqref{vphib}.
When $\eta_H>3$, the solution for 
$\bar{\eta}_H=3-\eta_H<0$ is
\begin{equation}
\label{hphid}
H(\phi)=M\sin\left(\sqrt{-\frac{\bar{\eta}_{H}}{2}}\,\phi\right),
\end{equation}
and the potential is
\begin{equation}
\label{vphid}
V(\phi)=M^2\left[3\sin^2\left(\sqrt{-\frac{\bar{\eta}_H}{2}}\phi\right)+\bar{\eta}_H\cos^2\left(\sqrt{-\frac{\bar{\eta}_H}{2}}\,\phi\right)\right]. 
\end{equation}
In the small field approximation with $|\phi|\ll 1$,
we get
\begin{equation}
\label{vphidapprox}
V(\phi)\approx M^2\left[\bar{\eta}_H-\bar{\eta}_H(3-\bar{\eta}_H)\frac{\phi^2}{2}\right]. 
\end{equation}
Due to the presence of the first term $\bar{\eta}_H$ in the right hand side of Eq. \eqref{vphidapprox}, 
the duality between $\eta_H>3$ and $\bar{\eta}_H=3-\eta_H$ 
does not hold for the solutions \eqref{hphid} and \eqref{vphidapprox} either even in the small field approximation.

For the particular solution
\begin{equation}
\label{hphia}
H(\phi)=M\cosh\left(\sqrt{\frac{\eta_{H}}{2}}\phi\right),
\end{equation}
the potential is
\begin{equation}
\label{vphia}
V(\phi)=M^2\left[3\cosh^2\left(\sqrt{\frac{\eta_H}{2}}\phi\right)-\eta_H\sinh^2\left(\sqrt{\frac{\eta_H}{2}}\phi\right)\right]. 
\end{equation}
In the small field approximation with $|\phi|\ll 1$, we get
\begin{equation}
\label{vphiaapprox}
V(\phi)\approx M^2\left[3+\frac{1}{2}\eta_H(3-\eta_H)\phi^2\right]. 
\end{equation}
It is interesting to note that in the small field approximation with $|\phi|\ll 1$, 
the potential \eqref{vphiaapprox} keeps to be the same if we change $\eta_H$ ($0<\eta_H<3$) to $\bar{\eta}_H=3-\eta_H$,
i.e., there is a duality between $\eta_H$ and $\bar{\eta}_H$ in the potential \eqref{vphiaapprox}, but the duality does not exist in $H(\phi)$. 
If $\eta_H>3$, then $\bar{\eta}_H=3-\eta_H<0$, the solution for $\bar{\eta}_H$ is
\begin{equation}
\label{hphic}
H(\phi)=M\cos\left(\sqrt{-\frac{\bar{\eta}_{H}}{2}}\phi\right),
\end{equation}
and the potential is
\begin{equation}
\label{vphic}
V(\phi)=M^2\left[3\cos^2\left(\sqrt{-\frac{\bar{\eta}_H}{2}}\phi\right)+\bar{\eta}_H\sin^2\left(\sqrt{-\frac{\bar{\eta}_H}{2}}\phi\right)\right]. 
\end{equation}
In the small field approximation with $|\phi|\ll 1$,
we get
\begin{equation}
\label{vphicapprox}
V(\phi)\approx M^2\left[3+\frac{1}{2}\bar{\eta}_H(3-\bar{\eta}_H)\phi^2\right]. 
\end{equation}
Therefore, in the small field approximation with $|\phi|\ll 1$,
the potential \eqref{vphicapprox} has the same form as \eqref{vphiaapprox} except that here $\bar{\eta}_H<0$, 
and it seems that there exists the duality between $\eta_H$ and $\bar{\eta}_H=3-\eta_H$ in the potential \eqref{vphicapprox}. 
When $\bar{\eta}_H<0$, $\eta_H=3-\bar{\eta}_H>0$, 
actually the potential \eqref{vphic} with $\bar{\eta}_H<0$ should be replaced by the potential \eqref{vphia} with $\eta_H>0$, 
so there is no duality between $\eta_H$ and $\bar{\eta}_H=3-\eta_H$ in the potential \eqref{vphic}
in the small field approximation.
In fact, if we replace $\eta_H=3+\alpha$ with $\alpha>0$ by  $\bar{\eta}_H=-\alpha$, the solutions \eqref{hphia} and \eqref{vphia} should be replaced by the solutions \eqref{hphic} and \eqref{vphic}, 
so the potential \eqref{vphiaapprox}  is dual to the potential \eqref{vphicapprox} under the interchange  between $\eta_H>3$ and $\bar{\eta}_H=3-\eta_H<0$.

Since logarithmic duality of the primordial curvature perturbation was found for the quadratic potential
in Ref.
\cite{Pi:2022ysn},
we will discuss whether the logarithmic duality exists in the constant-roll inflation for the particular solutions \eqref{vphiaapprox} and \eqref{vphicapprox}.
If $\epsilon_{H}\ll 1$ and $\phi\ll 1$, then substituting the potential \eqref{vphiaapprox} into Eq. \eqref{Vphi}, we get
\begin{equation}
\label{hphiapprox1} 
H^2(\phi)\approx V(\phi)/3\approx M^{2}.
\end{equation}
Plugging the potential \eqref{vphiaapprox} into Eq. \eqref{eomphi} and neglecting $\epsilon_H$, we get
\begin{equation}
\label{eomphi3}
\frac{d^{2}\phi}{dN^{2}}-3\frac{d\phi}{dN}+\eta_{H}(3-\eta_{H})\phi=0.
\end{equation}
The solution is
\begin{equation}
\label{phiN}
\phi(N)=C_{+}e^{\lambda_{+}(N-N_{e})}+C_{-}e^{\lambda_{-}(N-N_{e})},
\end{equation}
where $\lambda_{+}$ and $\lambda_{-}$ are 
\begin{equation}
\label{coef1}
\begin{split}
\lambda_{+}=\frac{3+\sqrt{9-4\eta_{H}(3-\eta_{H})}}{2},\\
\lambda_{-}=\frac{3-\sqrt{9-4\eta_{H}(3-\eta_{H})}}{2}, 
\end{split}
\end{equation}
and $N_e$ is the number of e-folds at the end of inflation.
Note that both $\lambda_+$ and $\lambda_-$ are invariant under the interchange between $\eta_H$ and $\bar{\eta}_H$,
and the inflationary model is not constant-roll inflation.
Remember that for constant-roll inflation with negligible $\epsilon_H$, 
the equation of motion for the scalar field satisfies Eq. \eqref{EOMN1}.
Combining Eqs. \eqref{eomphi3} and \eqref{EOMN1}, we get
\begin{gather}
\frac{d\phi}{dN}=\eta_H \phi,\\
\frac{d^2\phi}{dN^2}=\eta^2_H \phi.
\end{gather}
So the solution \eqref{phiN} is not the solution to constant-roll inflation, 
but we can still take $\eta_H$ as a parameter even though it loses the meaning of the second slow-roll parameter as defined in Eq. \eqref{etaconstant1}
and think that the quadratic potentials \eqref{vphiaapprox} and \eqref{vphicapprox} are motivated from constant-roll inflation.
Now back to the solution \eqref{phiN}, to determine the coefficients $C_{+}$ and $C_{-}$, we use the velocity of the scalar field
\begin{equation}
\label{piN}
\pi=-\frac{d\phi}{dN}=-C_{+}\lambda_{+}e^{\lambda_{+}(N-N_{e})}-C_{-}\lambda_{-}e^{\lambda_{-}(N-N_{e})},
\end{equation}
to get
\begin{equation}
C_{+}=-\frac{\lambda_{-}\phi_{e}+\pi_{e}}{\lambda_{+}-\lambda_{-}},\quad
C_{-}=\frac{\lambda_{+}\phi_{e}+\pi_{e}}{\lambda_{+}-\lambda_{-}},
\end{equation}
where $\pi_{e}$ is the value of $\pi(N)$ at $N_{e}$. 
Since $\lambda_+$ and $\lambda_-$ have the duality between $\eta_H$ and $\bar{\eta}_H$, so $C_{+}$ and $C_{-}$ also have the duality between $\eta_H$ and $\bar{\eta}_H$.
If we swap $\lambda_+$ and $\lambda_-$, then $C_+$ and $C_-$ are interchanged, and $\phi(N)$ and $\pi(N)$ are invariant.
Therefore
\begin{equation}
\label{expN}
\begin{split}
e^{\lambda_{+}(N-N_{e})}=\frac{\lambda_{-}\phi+\pi}{\lambda_{-}\phi_{e}+\pi_{e}},\\
e^{\lambda_{-}(N-N_{e})}=\frac{\lambda_{+}\phi+\pi}{\lambda_{+}\phi_{e}+\pi_{e}},
\end{split}
\end{equation}
and
\begin{equation}
\begin{split}
N-N_{e}&=\frac{1}{\lambda_{+}}\ln\bigg(\frac{\lambda_{-}\phi+\pi}{\lambda_{-}\phi_{e}+\pi_{e}}\bigg)\\
&=\frac{1}{\lambda_{-}}\ln\bigg(\frac{\lambda_{+}\phi+\pi}{\lambda_{+}\phi_{e}+\pi_{e}}\bigg).
\end{split}
\end{equation}

Using the $\delta N$ formalism \cite{Sasaki:1995aw, Wands:2000dp, Lyth:2004gb, Sugiyama:2012tj}, 
we get the primordial curvature perturbation
\begin{equation}
\label{zeta}
\begin{split}
\zeta&=\delta(N-N_{e})\\
&=\frac{1}{\lambda_{+}}\ln\bigg(1+\frac{\lambda_{-}\delta\phi+\delta\pi}{\lambda_{-}\phi+\pi}\bigg)-\frac{1}{\lambda_{+}}\ln\bigg(1+\frac{\delta\pi_{e}}{\lambda_{-}\phi_{e}+\pi_{e}}\bigg)\\
&=\frac{1}{\lambda_{-}}\ln\bigg(1+\frac{\lambda_{+}\delta\phi+\delta\pi}{\lambda_{+}\phi+\pi}\bigg)-\frac{1}{\lambda_{-}}\ln\bigg(1+\frac{\delta\pi_{e}}{\lambda_{+}\phi_{e}+\pi_{e}}\bigg).
\end{split}
\end{equation}

It can be seen that under the interchange between $\lambda_{+}$ and $\lambda_{-}$,
the two formulae for the primordial curvature perturbation are equivalent and the equivalence is guaranteed by taking the perturbation on Eq. \eqref{expN},
\begin{equation}
\label{deltapi1}
\bigg(1+\frac{\delta\pi_{e}}{\pi_{e}+\lambda_{+}\phi_{e}}\bigg)^{-\lambda_{+}}\bigg(1+\frac{\delta\pi_{e}}{\pi_{e}+\lambda_{-}\phi_{e}}\bigg)^{\lambda_{-}}=\bigg(1+\frac{\delta\pi+\lambda_{-}\delta\phi}{\pi+\lambda_{-}\phi}\bigg)^{\lambda_{-}}\bigg(1+\frac{\delta\pi+\lambda_{+}\delta\phi}{\pi+\lambda_{+}\phi}\bigg)^{-\lambda_{+}}.
\end{equation}
The formula \eqref{deltapi1} connects $\delta\pi_{e}$ with $\delta\phi$ and $\delta\pi$. 
In the same way, we can get the same result for the potential \eqref{vphicapprox}.
The result is the same as that found in \cite{Pi:2022ysn} and is called logarithmic duality. 
Therefore, there exists the logarithmic duality
and the duality between large and small $\eta_{H}$ for the primordial curvature perturbation in inflationary models with the potential \eqref{vphiaapprox} and \eqref{vphicapprox}.

\section{Conclusion}
\label{sec4}
PBHs and accompanying SIGWs are usually generated in inflationary models with inflection point.
Near the inflection point, the slow-roll parameter $\eta_H\approx 3$ and the inflation is also USR inflation which is a special case of the constant-roll inflation. 
For the constant-roll inflation, 
it is interesting that there is a duality between large and small $\eta_{H}$ in the observables $n_{s}$ and $r$ if the slow-roll parameter $\epsilon_{H}$ is negligible.
The duality between $\eta_H$ and $\bar{\eta}_{H}=3-\eta_H$ connects the constant slow-roll inflation with $|\eta_H|\ll 1$ and the USR inflation with $\eta_H\approx 3$,
so the behavior of primordial curvature perturbations in USR inflation can be understood from the usual slow-roll inflation.
However, the duality does not hold for the background evolution of the inflaton in general. 
If we neglect the contribution of $\epsilon_H$ in the background evolution of the inflaton,
we find that the inflaton potential has the duality between $\eta_H$ and $\bar{\eta}_{H}$.
We also derive the condition for neglecting $\epsilon_H$.
In some particular solutions of constant-roll inflation and in the small field approximation,
the inflaton potential takes a quadratic form and it remains the same under the interchange of the parameter $\eta_{H}$ and $\bar{\eta}_{H}$. 
When $0<\eta_{H}<3$, in the small field approximation, 
there is a duality between $\eta_{H}$ and $\bar{\eta}_{H}$ in the potential \eqref{vphiaapprox}, 
but there does not exist the duality in $H(\phi)$. 
When $\eta_{H}>3$ and $\bar{\eta}_{H}=3-\eta_H<0$, if we replace $\eta_{H}$ by $\bar{\eta}_{H}$, 
then the potential \eqref{vphiaapprox} for $\eta_{H}>3$ should be replaced by the potential \eqref{vphicapprox} for $\bar{\eta}_{H}<0$. 
Even in the small field approximation, both the potentials \eqref{vphiaapprox} and \eqref{vphicapprox} take the same quadratic form and are invariant under the interchange between $\eta_{H}$ and $\bar{\eta}_{H}$,
there does not exist the duality between $\eta_{H}$ and $\bar{\eta}_{H}$ in the potentials \eqref{vphiaapprox} and \eqref{vphicapprox}.

If we start with the quadratic potential \eqref{vphiaapprox} and \eqref{vphicapprox} without the restriction on constant-roll inflation, 
then using the $\delta N$ formalism we find that the primordial curvature perturbation exhibits a logarithmic duality if we neglect the contribution of $\epsilon_H$.
Since both $\lambda_{-}$ and $\lambda_{+}$ are invariant under the interchange between the parameter $\eta_{H}$ and $\bar{\eta}_{H}=3-\eta_H$, $\phi(N)$, $\pi(N)$ and the primordial curvature perturbation also exhibit the duality between the parameters $\eta_{H}$ and $\bar{\eta}_{H}$.

\begin{acknowledgments}
This research is supported in part by the National Key Research and Development Program of China under Grant No. 2020YFC2201504, the National Natural Science Foundation of China under Grant No. 12175184 and the Chongqing Natural Science Foundation
under Grant No. CSTB2022NSCQ-MSX1324.
\end{acknowledgments}

%

\end{document}